\documentclass[conference]{IEEEtran}
\IEEEoverridecommandlockouts
\usepackage[english]{babel}
\usepackage[utf8x]{inputenc}
\usepackage[T1]{fontenc}

\usepackage[colorinlistoftodos]{todonotes}
\usepackage[colorlinks=true, allcolors=blue]{hyperref}
\usepackage{float}

\usepackage{cite}
\usepackage{amsmath,amssymb,amsfonts}
\usepackage{algorithmic}
\usepackage{graphicx}
\usepackage{textcomp}
\usepackage{todonotes}
\usepackage{xcolor}
\usepackage{tikz}
\usetikzlibrary{shapes}
\usepackage{hyperref}
\PassOptionsToPackage{bookmarks={false}}{hyperref}
\usepackage[labelsep=quad,indention=10pt]{subfig}
\usetikzlibrary{calc}
\def\BibTeX{{\rm B\kern-.05em{\sc i\kern-.025em b}\kern-.08em
    T\kern-.1667em\lower.7ex\hbox{E}\kern-.125emX}}

\begin{document}
\title{A Public Network Trace of a Control and Automation System
}

\author{\IEEEauthorblockN{Gorby Kabasele Ndonda}
\IEEEauthorblockA{\textit{Department of Computer Science and Engineering}\\
\textit{UCLouvain, Belgium}\\
gorby.kabasele@uclouvain.be}
\and
\IEEEauthorblockN{ Ramin Sadre}
\IEEEauthorblockA{\textit{Department of Computer Science and Engineering} \\
\textit{UCLouvain, Belgium}\\
ramin.sadre@uclouvain.be}
}

\maketitle

\begin{abstract}
The increasing number of attacks against automation systems such as SCADA and their 
network infrastructure have demonstrated that there is a need to secure those 
systems. Unfortunately, directly applying existing ICT security mechanisms to automation
systems is hard due to constraints of the latter, such as availability requirements or 
limitations of the hardware. Thus, the solution privileged by researchers is the use of 
network-based intrusion detection systems (N-IDS).

One of the issue that many researchers encounter is how to validate and evaluate their 
N-IDS. Having access to a real and large automation systems for experimentation is 
almost impossible as companies are not inclined to give access to their systems due to 
obvious concerns. The few public traffic datasets that could be used for off-line 
experiments are either synthetic or collected at small testbeds. In this paper, we will 
describe and characterize a public traffic dataset collected at the HVAC management 
system of a university campus.

Although the dataset contains only packet headers, we believe that it can help 
researchers, in particular designers of flow-based IDS, to validate their solutions 
under more realistic conditions. 
The traces can be found on \url{https://github.com/gkabasele/HVAC_Traces}.

\end{abstract}


\section{Introduction}
Industrial control and automation systems, such as SCADA, have constraints different from
traditional enterprise IT systems, especially with regard to availability. For example, 
some end hosts, such as field device, have to operate continuously for several decades 
without being stopped. It is also not uncommon to see SCADA systems using outdated 
hardware and software as updating and patching is difficult and expensive. For this 
reason, network-based Intrusion Detection Systems (N-IDS) have been the privileged 
solution by researchers to secure SCADA systems\cite{Cheung:2007:SCADAIDS}
\cite{Colbert2016} since, in contrast to host-based IDS, they do not require 
modifications nor regular updates on the hosts. Among techniques for network-based 
intrusion detection, \emph{flow-based} methods have been employed by researchers for 
several years to build IDS for the Internet \cite{Sperotto2010} and also specifically for
SCADA systems \cite{Krejci2012,Barbosa2013, Zheng2017}.

One of the problems that researchers are facing is how to evaluate and 
validate their IDS solution. Getting access to real control and automation systems is 
hard because their owners are, for obvious reasons, rarely open to the idea
of running experiments in them. For this reason, most researchers rely either on 
experiments in testbeds or simulated environments or use traffic traces collected in such
environments for off-line experiments \cite{netresec2015,Lemay2016,Ring2017}. The 
advantage of these approaches is the reproducibility of the experiments, the 
flexibility to test different configurations, and the fact that no real system is 
put in danger. However, it requires that the used environments and traces 
accurately reflect the behavior of a real system, which is challenging to 
ensure. Publications relying on data from real systems for validation are much 
rarer and those that exist have not published their 
datasets \cite{Krejci2012,Zheng2017}, which makes it impossible to reproduce 
them by other researchers. To our knowledge, no public dataset from a real SCADA 
system or any other type of automation system exists so far.

In this paper, we present and describe a public Building Management System (BMS) 
dataset collected from part of the Heating, Ventilation, and Air Conditioning 
system (HVAC) of a university campus. A BMS is a type of automation system 
like SCADA. The dataset contains the headers of the network packets exchanged 
between field devices, control servers and human machine interfaces (HMI) of the 
BMS. The dataset does not contain the payload of the packets due to privacy and 
security reasons, but it can be used for purposes where traffic traces on packet 
header level or flow level are needed, such as
\begin{itemize}
\item flow-level traffic characterization,
\item parametrization of traffic models for synthetic traffic generation,
\item experimentation with realistic background traffic load.
\end{itemize}
To the best of our knowledge, the dataset only contains benign traffic. This 
means that the traces cannot be directly used to validate flow-based intrusion 
detection systems due to the absence of attacks (unless the goal is to evaluate 
false positives). However, researchers can combine them with malicious traffic or 
use them to generate synthetic traffic, as indicated above.

The structure of this paper is as follows: Section~\ref{sec:description} describes 
the system from which the trace has been collected. Section~\ref{sec:methodology} 
explains the  methodology used to collect the trace. Section~\ref{sec:results} 
presents important traffic characteristics and findings, followed by a 
discussion in Section~\ref{sec:discussion}. Finally, the paper concludes 
in Section~\ref{sec:conclusion}.
\section{Description}\label{sec:description}
In this section we describe the BMS from which the trace has been collected. 
Due to privacy and security reasons, we will not publish certain details, such 
as the deployed control software, information related to non-BMS services, or 
the physical configuration of the communication infrastructure.

\subsection{HVAC Management System}\label{subsec:hvac}
The trace contains the network communication of a HVAC management system. The 
system has been deployed by and is managed by Honeywell.

The system is fully automated, with a server communicating with peripheral 
devices deployed on the campus to control heating in offices, classrooms, and 
lecture halls. The part of the system that we have monitored controls around 15 
to 20 buildings. It has been optimized for energy efficiency: When a teacher books 
a classroom or lecture hall via the booking application, a message is sent to the 
main server so that it can instruct the PLC or RTU responsible for that room to 
activate the heating several minutes before people are coming in. Operators access 
the system through the Human Machine Interface (HMI), which consists of 
dedicated workstations with a graphical user interface, to see in real-time the
current system state or to set temperature goals manually.

\subsection{System overview}

The BMS, topology-wise, follows a typical design for automation systems \cite{nist-ics}. A communication network interconnects the control server, the HMI stations, and the peripheral devices. The sensors and actuators are attached to the latter. Therefore, the peripheral devices are similar to PLCs and RTUs in SCADA systems: They collect and send sensor data to the server and their behavior can be (manually) influenced from the HMIs, but most of the time they execute simple control programs with setpoints defined by the server. The network is isolated from the rest of the campus network. The communication protocols are proprietary and use TCP/IP as transport layer.

It is important to note that the peripheral devices are not directly connected to the network. Instead, a group of devices is connected to a gateway which acts as an end point for the TCP connection from the control server (Figure~\ref{fig:overview}). This allows to also use devices which are not IP-capable. In the part of the network that we consider for our dataset there are 8 such gateways, each one connected to 1 to 15 peripheral devices.

\newcommand{\shiftpoints}{4pt}

\begin{figure}[t]
\centering
\begin{tikzpicture}[
	shifttl/.style={shift={(-\shiftpoints,\shiftpoints)}},
    shifttr/.style={shift={(\shiftpoints,\shiftpoints)}},
    shiftbl/.style={shift={(-\shiftpoints,-\shiftpoints)}},
    shiftbr/.style={shift={(\shiftpoints,-\shiftpoints)}},
]
	\node [cloud, draw,cloud puffs=10,cloud puff arc=120, aspect=2, inner ysep=1em](network) {Network};
    \node[above=3cm of network]{System part monitored};
    \node[above left=of network] (hmi-a) {\includegraphics[scale=0.3]{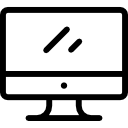}};
    \node[left = of hmi-a] (hmi-b) {\includegraphics[scale=0.3]{fig1b.png}};
    \node[above =1pt of hmi-b] (hmi-label){HMI};
    \node[above right=of network] (mtu) {\includegraphics[scale=0.075]{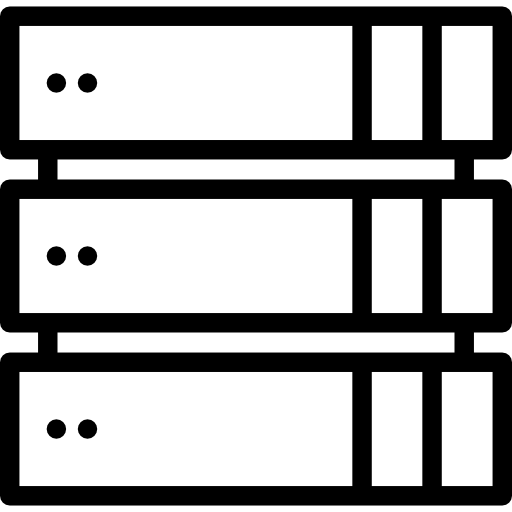}};
    \node[above =1pt of mtu] (mtu-label) {Control Server};
    \node[left=of network] (gw-a) {\includegraphics[scale=0.075]{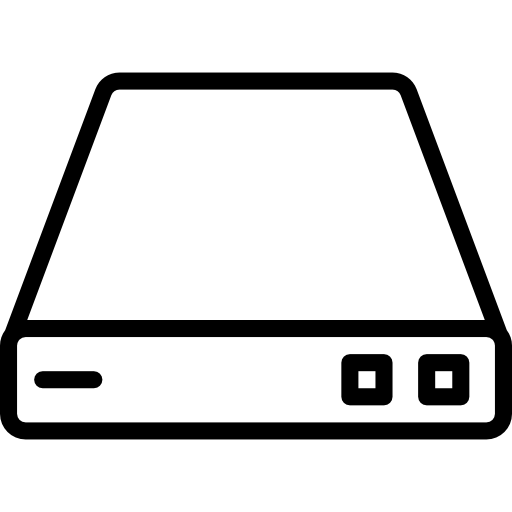}};
    \node[below=of gw-a] (gw-b) {\includegraphics[scale=0.075]{fig1d.png}};
    \node[below =of network] (gw-c) {\includegraphics[scale=0.075]{fig1d.png}};
    \node[right=of network](gw-d){\includegraphics[scale=0.075]{fig1d.png}};
    \node[above =1pt of gw-d] {Gateway};
    \node[below left=0.3cm of gw-a](gw-e){\includegraphics[scale=0.075]{fig1d.png}};
    \node[below left=0.5cm of gw-c](plc-a) {\includegraphics[scale=0.7]{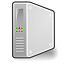}};
    \node[left=1pt of plc-a]{Peripheral devices};
    \node[below right=0.5cm of gw-c](plc-b){\includegraphics[scale=0.7]{fig1c.png}};
    \node[below=0.5cm of plc-a](valve-a){\includegraphics[scale=0.3]{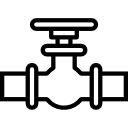}};
    \node[left=1pt of valve-a]{Sensors and actuators};
    \node[below=0.5cm of plc-b](valve-b){\includegraphics[scale=0.3]{fig1e.png}};
    
    \draw[-] (hmi-a) -- (network);
    \draw[-] (hmi-b) -- (network);
    \draw[-] (mtu) -- (network);
    \draw[-] (gw-a) -- (network);
    \draw[-] (gw-b) -- (network);
    \draw[-] (gw-c) -- (network);
    \draw[-] (gw-d) -- (network);
    \draw[-] (gw-e) -- (network);
    \draw[-] (gw-c) -- (plc-a);
    \draw[-] (gw-c) -- (plc-b);
    \draw[-] (plc-a) -- (valve-a);
    \draw[-] (plc-b) -- (valve-b);
    
    \begin{scope}[transform shape] 
	\path [draw,rounded corners]  
               ([shifttl] hmi-label.north west) 
            -- ([shifttr] mtu-label.north east) 
            -- ([shiftbr] gw-d.south east)
            -- ([shiftbr] valve-b.south east) 
            -- ([shiftbl] valve-a.south west) 
            -- ([shiftbl] network.west) 
            -- ([shiftbl] hmi-b.south west) 
            -- cycle; 
\end{scope}

\end{tikzpicture}
\caption{Network overview}
    \label{fig:overview}
\end{figure}
\section{Methodology}\label{sec:methodology}

In this section, we describe the methodology we used to collect the BMS traffic as well as the post processing applied to obtain the dataset. 

\subsection{Packet collection}\label{sub:collecting}

In order to collect all network communication in the BMS network, we capture packets at two routers via port mirroring. The usage of two vantage points is necessary because the paths between the server and the peripheral devices are asymmetric.

We use \emph{tcpdump} to capture the packets. For each packet, 
\emph{tcpdump} records a timestamp of when the packet was captured based on the clock of the capturing device. Since we use two devices, we synchronize their clocks with NTP so that packets appear later in the correct order when merging the individual packet traces based on the timestamps. Unfortunately, it showed in our first test runs that this clock synchronization is not sufficient and time shifts ranging from 3ms to 10ms can be observed, resulting in wrong packet orders.

To obtain a more accurate timing, we periodically compute the shift between the two clocks and correct the recorded timestamps accordingly when merging the traces. To this end, capture device $A$ sends every second an ICMP echo request to device $B$ which will reply with an ICMP echo reply. These ICMP packets appear in the traces of both devices. The current time shift $s$ between the two clocks is then estimated by 
  $$s =   t_B - t_A - RTT/2 $$
where $t_i$ is the timestamp of the ICMP echo request as seen by device $i$. The round-trip time $RTT$ between the two devices is estimated as
  $RTT = t'_A - t_A$
where $t'_A$ is the timestamp of the ICMP echo reply as seen by device $A$. Computing the network delay between the two devices in this way assumes that the delay is identical in both directions. 

Using this approach, the number of out-of-order packets in a trace of one hour duration can be reduced from 8500 to 85. By increasing the frequency of the ICMP packets, we could further decrease this number but we do not want to send too many packets and add artificial load to the network.   

\subsection{Merging and anonymization process}

When merging the packet traces from the two capture devices, we calculate for 
every ICMP echo request the time shift $s$ using the method described above. We 
then adjust the timestamps of all packets captured by device $A$ in the second 
following the ICMP request, i.e. the period of time until the next ICMP request.
At the end, we obtain merged and time-corrected traces containing the packets captured by
the two devices. In a filtering step, we only keep packets that either originate from or are destined to an host being part of the HVAC system, typically the control server, an HMI, or a gateway. We remove packets from protocols that are purely related to network management, namely ARP, STP, VRRP, LLMR. The original traces also contain SNMP traffic sent by the workstations to several devices inside and outside the network. Unfortunately, only the SNMP requests are visible to our capture devices so we remove them. We publish the trace in form of pcap files where each pcap file contains one hour of traffic.

The owner of the BMS has demanded that the dataset is anonymized before publication.
To do so, several steps were taken. First, we use the CryptoPan tool \cite{FAN2004253} to 
anonymize the IP addresses in a prefix-preserving manner. This tool uses cryptographic techniques and has the advantage of being consistent across several files given that the same key is used for the anonymization. We then truncate the packet payload to only keep 
the headers and zero out the Organizationally Unique Identifier (OUI) of the MAC address.

\subsection{Ethical considerations}

Before publishing the traces, we discussed with the parties that would be affected by having the traces publicly available, following the advices in \cite{menlo}. These parties included the Chief Information Security Officer of the university the dataset was collected at and Honeywell, the company who deployed the HVAC system and who are the manufacturers of the devices and the authors of the protocols used by them. We got their agreement after we explained what we wanted to publish and why.

During the processing of the traces, we made sure that no personal data was collected and
that the traces can not be used to infer the behavior of individual people. The dataset 
only contains network traffic generated automatically, meaning that it is not the direct 
result of human action, except for some part of the communication between the HMI 
workstations and the control server.
\section{Traffic characteristics}\label{sec:results}

This section presents the main characteristics of the collected traces. We will discuss 
some of these characteristics in terms of unidirectional flows. We define a flow using 
the usual five-tuple (source IP address, destination IP address, source port, 
destination port, transport protocol). When we refer to flow size or packet size in the 
following, we always mean the size of the payload, excluding headers. For TCP flows, we 
only consider the ones where at least one byte of payload was exchanged, thus removing 
connections which failed to be established. Note that our flow definition ignores TCP 
flags (SYN, FIN etc.), i.e. a flow more or less corresponds to the definition of a flow 
record in Netflow/IPFIX\cite{RFC7011} with active and inactive timeout set to infinity.


\subsection{Overview}

We collected the traces over a period of one week (7 days) in order to capture the 
behavior of the BMS on weekdays as well as during the weekend. The capture started a 
Friday afternoon at approximately 2PM.
Table~\ref{tab:system-sum} shows a summary of the part of the HVAC system we have 
monitored. The right three columns give basic statistics of the bidirectional 
communication between HMI$\leftrightarrow$server and server$\leftrightarrow$gateways, 
respectively. The peripheral devices are not IP-capable, so we could not capture the 
traffic between them and the gateways.

\begin{table}[h]
\centering
	\begin{tabular}{|c|c||c|c|c|c|}
   	\hline
       Host Type&Number&Communicate&Payload&Nbr pkts\\
       &&with&exchanged&exchanged\\
       \hline
       HMI stations&7&Control Server&6.7GB&24.7M\\
       Control Server&1&Gateways&387.7MB& 11M\\
       Gateways&8&Periph. devices&--&--\\
       Periph. devices&58&--&--&--\\
		\hline
	\end{tabular}
   \caption{HVAC summary}
   \label{tab:system-sum}
\end{table}

In total, we collected 11.2~GB of packet data (corresponding to 4GB of packet headers) 
from 47 IP addresses and approximately 23,000 flows. Table~\ref{tab:trace-sum} shows 
some characteristics of the entire dataset and the flows. As expected of automation 
network, flows are small, the largest one transported 614.7 MBytes over one week.

\begin{table}[h]
\centering
	\begin{tabular}{|c|c||c|c|c|c|}
   	\hline
       \multicolumn{2}{|c||}{dataset} & \multicolumn{3}{c|}{flows}\\
       \multicolumn{2}{|c||}{} & \multicolumn{1}{c}{} & bytes & packets\\
       \hline
       size&11.2GB&min&5B&1\\
       duration&7 days&avg&342.2kB&1725.08\\
       Nbr flows&23K&max&614.7MB&5.1M\\
       Nbr IP&47&&&\\
		\hline
	\end{tabular}
   \caption{Trace Summary}
   \label{tab:trace-sum}
\end{table}

\begin{figure*}[h]
	\centering
	\captionsetup[subfigure]{oneside,margin={0.5cm,0cm}}
	\subfloat[CDF of the number of packets per flow]{
    	\includegraphics[scale=0.36]{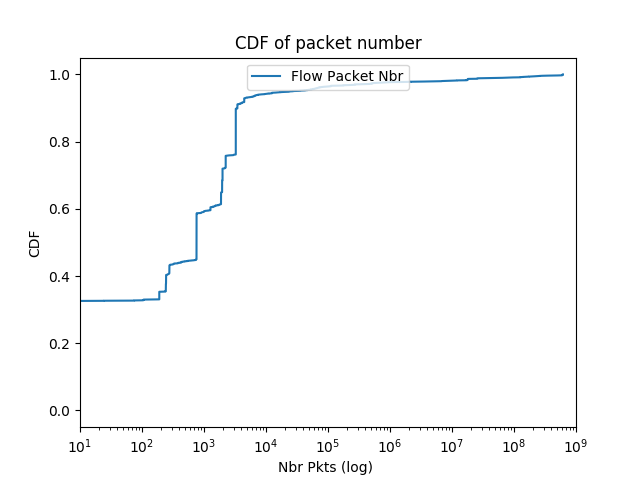}
        \label{subfig:pkt-cdf}
    }
    \subfloat[CDF of the number of bytes (in kBytes) per flow]{
    	\includegraphics[scale=0.36]{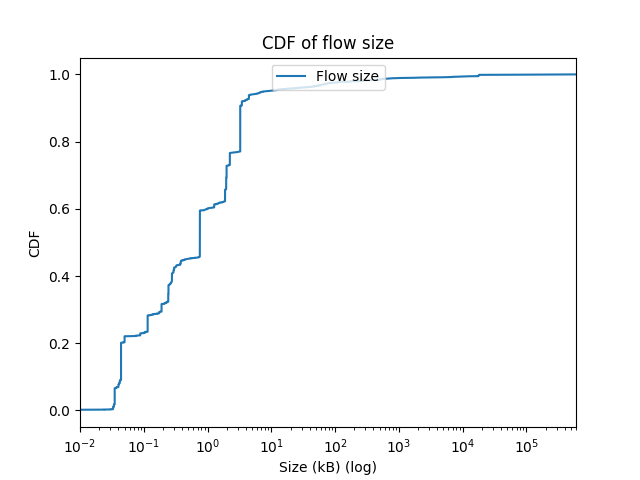}
        \label{subfig:pkt-size}
    }
     \subfloat[CDF of the flow duration (in hours)]{
    	\includegraphics[scale=0.36]{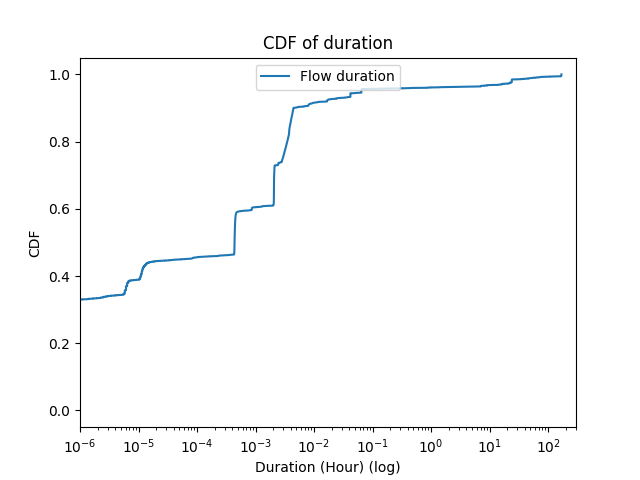}
        \label{subfig:cdf-duration}
    }
\caption{Flow characteristics}
\vspace{-6pt}
\label{fig:cdf}
\end{figure*}

Figure~\ref{fig:cdf} shows the empirical CDF of (a) the number of packets per flows, (b)
the number of bytes per flow, and (c) the flow duration. The CDFs show that there is a 
significant number of flows with a very long duration and a large number of packets, 
although the distribution of the number of bytes per flow is rather conservative. Such a
behavior is typical for automation systems where the server and the peripheral devices 
periodically exchange small control packets through more or less permanent TCP 
connections. In fact, all those connections in our trace lasted over the entire 
measurement period.

\subsection{Application protocols}

Most of the observed traffic is from communication between the HMI stations and 
the control server. The HMI stations and the control server communicate through 
several port. First, they exchange Remote Procedure Calls for Distributed 
Computing Environment (DCE/RPC) packet \cite{dcerpc}. The control server acts as 
the DCE/RPC server on port 135. Surprisingly, an inspection of the packets revealed 
that other ports were used as well to exchange DCE/RPC packets. Those ports were 
not fixed; they changed during the duration of the capture.
The protocol used for the communication between the server and the gateways is 
proprietary (port 2499). 

In addition to the protocols used between the different components of the HVAC system, 
we observed other protocols. These protocols include ones from the NetBIOS family (NBDS,
NBNS, NBSS) and are used by the HMI workstations and the control server to exchange 
data with each other and discover available services in the network.

We also observed some S7 communication, a protocol used by Siemens PLCs. This
traffic was related to another part of the BMS but we kept it in the dataset as
it demonstrates an example of heterogeneity in automation systems. Web traffic was
also observed, we discuss it in Section~\ref{subsec:updates}.

\subsection{Network activity time series}

As mentioned in Section~\ref{subsec:hvac}, the peripheral devices are connected to 
the IP network through gateways. The various devices differ in age and capabilities
and therefore depict different communication characteristics. We illustrate this in
the following with two gateways $X$ and $Y$ which connect 11, resp. 13, 
peripheral devices. In the part of the BMS monitored by us, 6 gateways have the
same profile as gateway $X$ and 2 gateways are similar to gateway $Y$, but 
with different numbers of peripheral devices connected to them.

\begin{figure*}[h]
  \vspace{-20pt}
	\centering
	\subfloat[Packets per hour]{
    	\includegraphics[scale=0.36]{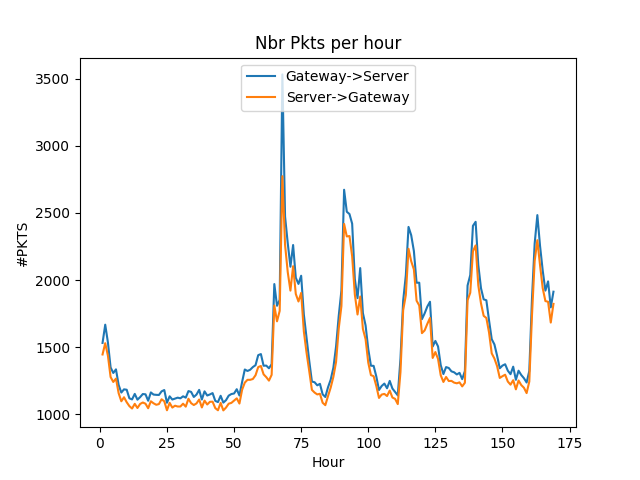}
        \label{subfig:pkt-a}
        }
    \subfloat[KBytes per hour]{
    	\includegraphics[scale=0.36]{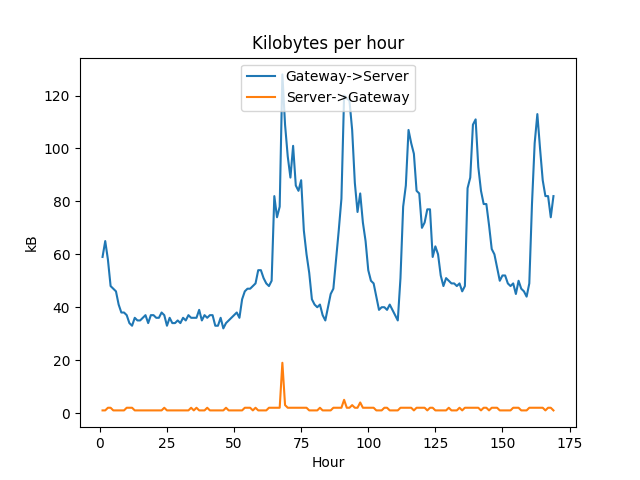}
        \label{subfig:size-a}
        }
    \subfloat[Packet size distribution]{
    	\includegraphics[scale=0.36]{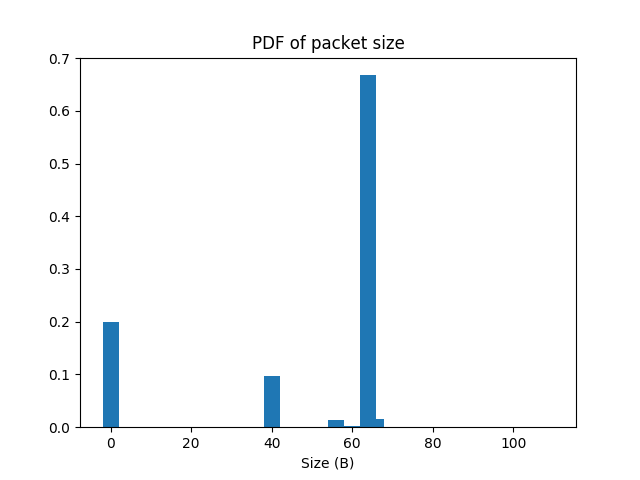}
        \label{subfig:pdf-size-a}
    }
\caption{Traffic observed at gateway $X$}
\vspace{-5pt}
\label{fig:gwa}
\end{figure*}

\begin{figure*}[h]
  	\vspace{-15pt}
	\centering
    \subfloat[Packets per hour]{
    	\includegraphics[scale=0.36]{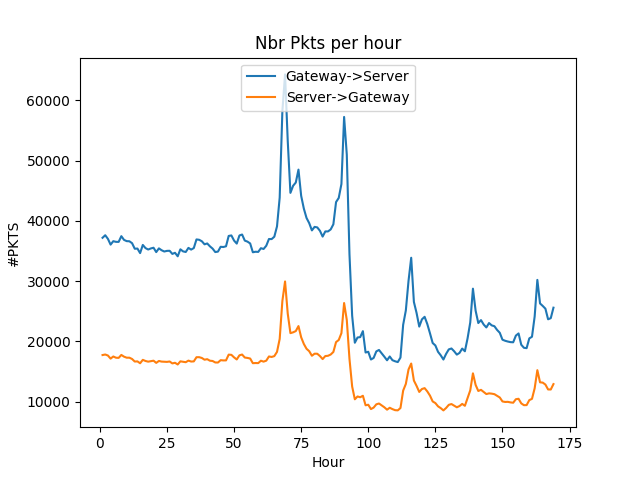}
        \label{subfig:size-b}
    }
    \subfloat[KBytes per hour]{
    	\includegraphics[scale=0.36]{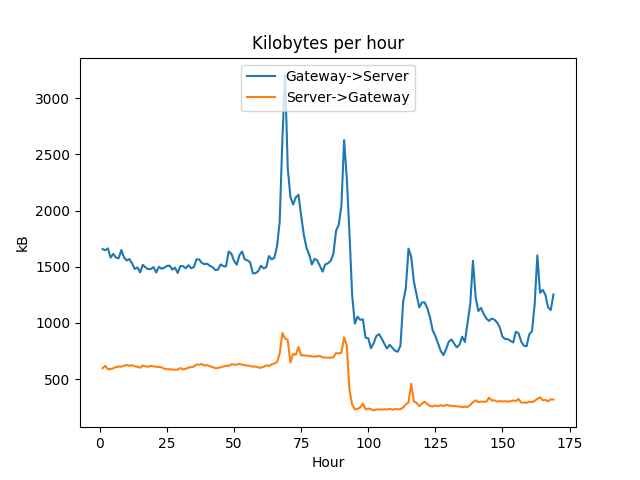}
        \label{subfig:pkt-b}
    }
    \subfloat[Packet size distribution]{
    	\includegraphics[scale=0.36]{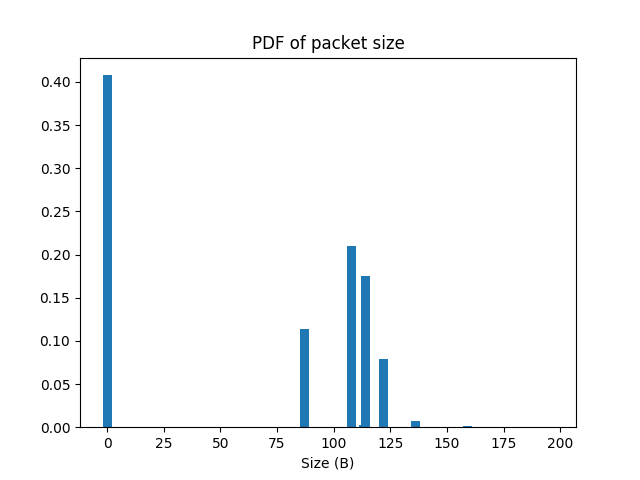}
        \label{subfig:pdf-size-b}
    }   
\caption{Traffic observed at gateway $Y$}
\label{fig:gwb}
\end{figure*}

We plot the time series of the number of packets and number of bytes exchanged 
per hour between the main server and the 
gateways. Figures~\ref{fig:gwa}a and~\ref{fig:gwa}b 
(resp. \ref{fig:gwb}a and \ref{fig:gwb}b) show these plots for gateway $X$ (resp. $Y$).
We note some differences between the two gateways. First, despite 
having a comparable number of devices attached to them, the number of 
packets observed at gateway $Y$ is considerably higher than at gateway $X$. A 
similar observation can be made for the number of exchanged bytes.
We also notice how differently the packets are generated in both directions. 
In gateway $X$, the number of packets sent to the server is close to the number 
of packet sent in the other direction. In fact, most of the packets sent by the 
server to gateway $X$ are ACK segments. We can see that the server behaves 
differently in the case of gateway $Y$: It actively sends segments with control 
commands to the gateway. Since the gateways are connected to devices of 
different models, it means that the devices connected at gateway $Y$ are 
more data intensive.

Finally, we observe diurnal and weekly patterns in the timeseries. The measurement
was started on a Friday at around 2PM, therefore the first 55 hours of low activity
in the timeseries correspond to the weekend. This behavior is understandable
when considering the nature of the physical  processes controlled by the BMS, but
it should be noted that not all automation systems behave necessarily like 
this \cite{Barbosa2012}. Interestingly, gateway $Y$ shows lower activity during 
the night than during the weekend although the buildings are not used in 
both situations. Our assumption is that there is a kind of sleep mode enabled 
during the night which decreases the frequency at which data are exchanged.

\subsection{Packet size distribution}

SCADA network protocols such as Modbus/TCP\cite{Modbus} or DNP3 require that the 
payload fits in a single IP packet to avoid packet fragmentation. These 
protocols aim at low latencies or even real-time usage, so packet fragmentation
is undesired. We study the packet sizes of for the entirety of the packets sent
from both  gateways to the server. The gateway $X$ sent 255,251 packets and the
gateway $Y$ sent 5,183,021 packets. 
Figure~\ref{subfig:pdf-size-a} and~\ref{subfig:pdf-size-b} show the 
normalized histograms of the observed packet payload sizes in bytes for 
each gateway. Similar to SCADA systems, the packet sizes are rather small 
in both cases --- they do not exceed 170 bytes. Moreover, only a few different 
sizes can be observed. Similar to other characteristic, there is a difference
between the two gateways. Most of the packet sent by the gateway $X$ contain
actual data and there is only a small percentage (approx. 20\%) of empty ACK
while gateway $Y$ sends more that 2 million empty packets.

\subsection{Flow stability}\label{subsec:updates}

When comparing traditional ICT networks and automation networks, it is a commonly 
accepted fact that the latter are more stable in term of traffic activities, such as the
number of TCP connections established. This is also mostly true for the 
BMS network studied in this paper.

\begin{figure}[h!]
	\centering
    \includegraphics[scale=0.45]{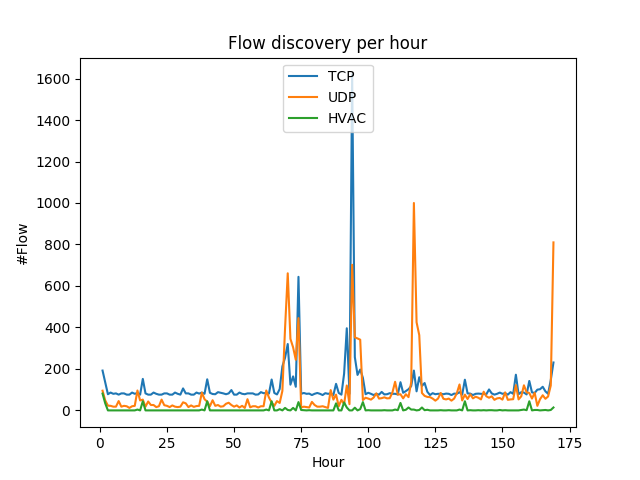}
    \caption{Flows seen per hour}
    \label{fig:new-flow}
\end{figure}

We show in Figure~\ref{fig:new-flow} the number of new flows discovered per 
hour. For this figure, we have counted flows directly related to the HVAC 
system, i.e. flows between the control server, the HMI workstations and the 
gateways, separately from other UDP and TCP flows. As it can be seen in 
the figure, except for the first hour when we start the capture, the number 
of flows discovered for the HVAC system is mostly constant. Actually, flows 
from the gateway to the control server lasted for the whole duration of the 
capture and there were no new flow created. The peaks that can be seen on the 
HVAC result from the change of the port used by the control server for 
DCE/RPC. The number of TCP flows discovered usually does not exceed 500 except 
at some point where we can see high peak. The TCP and UDP flow, apart form the 
HVAC system, are mostly HTTP/HTTPS and DNS flows created when the HMI stations 
check if there are new updates. The peaks are caused by actual software updates 
that took place during the measurement period. We have deliberately kept 
these "anomalies" in the dataset as they illustrate that even stable BMS 
systems can show irregular traffic patterns which might confuse intrusion 
detection systems.

\section{Discussion and related work}\label{sec:discussion}

The characteristics of network traffic in industrial control and automation systems have been discussed in several publications. The SCADA network traffic of different utility companies is studied in \cite{Barbosa2012,Barbosa2012b}. In \cite{Krejci2012} and \cite{Zheng2017}, the authors analyze the network traffic of BACnet-based BMS. To our knowledge, none of the datasets used in those and similar publications is publicly available due to privacy and security concerns. Therefore, researchers without access to a real system have to rely on public datasets collected in testbeds and virtual environments \cite{netresec2015,Lemay2016} or setup their own infrastructure for experiments, for example in \cite{Lin2013}.

As explained, our dataset does not contain payload data nor malicious traffic. Nevertheless, it can be used in research on flow-based network intrusion detection systems that do not rely on deep packet analysis. With tools like tcpreplay \cite{tcpreplay}, researcher can replay the traces to obtain (benign) traffic that can be mixed with synthetic attack traffic for their experiments.
Moreover, the traces contain valuable packet-level information for anomaly-based intrusion detection \cite{GARCIATEODORO200918}, such as the time\-stamp and size of each packet. This data allows to study the dynamic of the HVAC flows which is useful to create a model of the system.

The traces can also serve as a basis for trace generation. As mentioned earlier, researchers experiment on traces collected from simulated environments or testbeds since they do no have access to real systems. To be of relevance, the simulated environments must behave as close as possible to a real systems, which is hard to ensure. We believe that the traces can help to improve the accuracy of simulated environments. There is some work on how to generate synthetic traces from another one \cite{DBLP:journals/corr/abs-1810-07795,varet14}. It relies on statistical models and machine learning techniques to generate traces that have similar properties to the one they were generate from. We will also work in that direction as future work.

\section{Conclusion}\label{sec:conclusion}

Doing research on industrial control and automation systems security is challenging 
because researchers usually do not have access to real systems for 
experimentation and existing public datasets for off-line experimentation are 
rare and stemming from testbeds or simulated environments.

In this paper, we present a public BMS network trace collected from the 
HVAC management system of a university campus. We show that the trace depicts 
the common characteristics of automation systems, such as long lived connections, 
but also follows a diurnal pattern similar to those caused by human activities 
in traditional IT networks.

To our knowledge, this is the first public trace from a real BMS. The traces can be found on \url{https://github.com/gkabasele/HVAC_Traces}.

\bibliography{references}
\bibliographystyle{plain}

\end{document}